\documentclass{article}

\usepackage{arxiv}

\usepackage[utf8]{inputenc}
\usepackage[T1]{fontenc}
\usepackage{hyperref}
\usepackage{url}
\usepackage{booktabs}
\usepackage{amsfonts}
\usepackage{nicefrac}
\usepackage{microtype}
\usepackage{cleveref}
\usepackage{graphicx}
\usepackage{natbib}
\usepackage{doi}
\usepackage{tikz}
\usepackage{csquotes}
\usetikzlibrary{arrows.meta, positioning, fit, calc}

\title{Verification-Conditioned Use: A Qualitative Study on How Generative AI Reshapes Learning, Autonomy, and Market Entry for Junior Software Developers}

\newif\ifuniqueAffiliation
\uniqueAffiliationtrue

\ifuniqueAffiliation
\author{ Pedro Henrique Andriotti Bastos \\
	Computer Science Program\\
	CESAR School\\
	Recife, Brazil \\
	\And
	Danilo Ribeiro \\
	Computer Science Program\\
	CESAR School\\
	Recife, Brazil \\
}
\fi

\renewcommand{\shorttitle}{Verification-Conditioned Use of Generative AI by Junior Developers}

\hypersetup{
pdftitle={Verification-Conditioned Use: A Qualitative Study on How Generative AI Reshapes Learning, Autonomy, and Market Entry for Junior Software Developers},
pdfsubject={cs.SE, cs.HC},
pdfauthor={Pedro Henrique Andriotti Bastos, Danilo Ribeiro},
pdfkeywords={Generative Artificial Intelligence, Junior Developers, Software Engineering Education, Thematic Analysis, Human-AI Interaction},
}

\begin{document}
\maketitle

\begin{abstract}
\textbf{Objective:} to investigate how the use of generative Artificial Intelligence (AI) tools affects the early stages of a career in software development, from the perspective of the newcomers themselves. \textbf{Method:} thirteen interns and junior developers were interviewed individually, by videoconference. Interviews were analyzed using the six phases of Braun and Clarke's thematic analysis, with inductive coding and a semantic approach. \textbf{Results:} sixteen themes emerged, organized around a central concept: \emph{verification-conditioned use}. Across the study's four research questions (usage patterns, learning, autonomy, and market entry), the criterion that most often decides between AI and manual work is not deadline or task complexity, but the ability to check the result. Two themes expose tensions in newcomers' self-perception: the \emph{autonomy paradox} (feeling more capable yet less in ownership of the result) and the first-person denial of dependence. Together, these findings point to a theoretical contribution, the \emph{formative paradox}: the shallow learning that AI induces makes it harder to build the very critical-judgment competence that, according to participants, the market has begun to demand. \textbf{Conclusion:} what makes AI use sustainable, from participants' own point of view, is not the tool itself but the individual practice of reviewing before accepting, refusing to use AI without understanding it, asking the tool for explanations, and keeping deliberate practice outside of AI-assisted work.
\end{abstract}


\section{Introduction}
\label{sec:intro}

Generative AI has entered the daily routine of software development in a few short years. Tools such as GitHub Copilot, Cursor, ChatGPT, Claude Code, and Codex are now part of the everyday practice of programmers. Tasks that once required searching documentation, browsing forums, or manual prototyping are now supported by assistants that generate code, explain concepts, and propose solutions from the context of a question \citep{Mah25,Hib25}.

This shift does not affect every profile equally. Interns and junior developers are still building their technical repertoire and looking for a place in the job market, and for this group AI use has become close to compulsory: in some teams, deadlines and delivery expectations already assume that AI is used, which can lead newcomers to adopt it before they have enough of a knowledge base to critically evaluate what it produces. Prior work reports productivity and perceived performance gains from AI use \citep{Szo25}, alongside preliminary evidence of side effects such as shallower learning, possible skill-retention loss, and passive usage patterns, particularly requests for a ``direct solution'' \citep{Mah25,Hib25}. This study investigates these effects specifically among newcomers to the market, rather than among established professionals.

\subsection{Research Problem}
\label{sec:problem}

Existing evidence already indicates that AI helps newcomers gain immediate productivity. This study addresses a less obvious question: in which aspects, and in what ways, does early AI adoption reconfigure the start of a career in software development. Three tensions structure the problem. First, \emph{immediate productivity versus depth of learning}: AI substantially reduces the time to complete tasks, but there are indications that this acceleration may compromise the depth with which newcomers build technical knowledge, with consequences for retention and the ability to solve complex problems unassisted \citep{Mah25}. Second, \emph{perceived autonomy versus actual dependence}: having a tool at hand that answers questions, generates code, and explains concepts at any moment increases the newcomer's sense of autonomy, but may also create dependence; when the tool is unavailable, or when the professional must act outside the scenarios where AI is effective, confidence and performance may falter \citep{Hib25}. Third, \emph{reconfiguration of market-expected competencies}: activities traditionally assigned to juniors, such as CRUD operations, scripts, and small refactors, are now efficiently handled by AI itself, which may raise the entry barrier and shift the expected skill set of a newcomer toward architecture, business-rule translation, and critical review of tool output.

Building on these tensions, the study investigates, from the professionals' own perspective, how AI affects entry into the software job market. The main research question is:

\begin{displayquote}
\emph{How does the adoption of generative AI tools by interns and junior developers impact their day-to-day work, learning process, task execution, and professional perception when entering the software development job market?}
\end{displayquote}

This question unfolds into four sub-questions, each defining a thematic axis that guided coding and theme aggregation:

\begin{itemize}
	\item \textbf{RQ1 -- Usage patterns and decision criteria}: What criteria do interns and junior developers use to decide between using AI and solving a task manually, and how do these criteria vary by activity type (architecture, business rules, CRUD, front-end, testing, refactoring)?
	\item \textbf{RQ2 -- Reconfiguration of the learning process}: How does AI use reconfigure learning for interns and juniors, with attention to depth of knowledge, retention, and the nature of the questions taken to more experienced professionals?
	\item \textbf{RQ3 -- Perceived autonomy, confidence, and dependence}: How do interns and juniors perceive the effects of AI use on autonomy, confidence, and dependence, and what self-regulation strategies do they report adopting?
	\item \textbf{RQ4 -- Reconfiguration of market entry}: How is the presence of AI perceived by newcomers as a factor reconfiguring expected competencies, the entry barrier, and the future of work in the software market?
\end{itemize}

These four research questions are not independent hypotheses to be individually confirmed or rejected. Following the logic of thematic analysis \citep{BC06}, they define the analytical lenses used during theme development: each sub-question delimits a thematic axis around which codes were grouped into candidate themes (analysis phase 3) and later consolidated into final themes (phase 5). A single interview excerpt can therefore inform more than one axis, and the themes that emerge under one research question are expected, and in fact turn out, to interact with themes under the others, which is precisely what the integrated synthesis in Section~\ref{sec:synthesis} makes explicit.

\subsection{Contributions}
\label{sec:contributions}

This study contributes: (i) a characterization, from newcomers' own accounts, of AI usage patterns and the decision criteria between tool use and manual resolution; (ii) a description of perceived transformations in the learning process and in the nature of questions taken to more experienced professionals; (iii) an account of how newcomers perceive the effects of AI on autonomy, confidence, and dependence, and which self-regulation strategies they adopt; and (iv) an understanding of how AI reconfigures, from newcomers' own perspective, expected competencies, the entry barrier, and the future of software work. The integration of these four axes around a single decision criterion, \emph{verification-conditioned use}, and the identification of a \emph{formative paradox} are the study's main theoretical contributions.

\section{Background and Related Work}
\label{sec:related}

\subsection{Generative AI in Software Development}
\label{sec:background-ai}

Large language models have changed the way software is written. Tools such as GitHub Copilot, Cursor, ChatGPT, Claude Code, and Codex have become part of developers' routines within a few years, with rapid, global-scale adoption \citep{Doh23,Szo25}. The literature describes this as a shift in role: AI stops being a one-off tool and starts acting as a co-producing agent alongside the developer \citep{Doh23,Szo25}, suggesting code contextually, completing functions, and proposing project scaffolding from natural-language prompts. The novelty of this generation of tools is not raw speed but the capacity to answer in a personalized way, tailoring the explanation to the specific doubt of whoever is asking \citep{Hib25,Mah25}, which is what makes these assistants compete directly with traditional learning sources such as documentation and Stack Overflow.

This matters disproportionately for interns and junior developers. Unlike established professionals, newcomers are simultaneously using AI to produce work \emph{and} to build the technical repertoire they do not yet have; every AI-mediated shortcut is, for them, also a possible shortcut around a learning opportunity. Established developers can treat AI primarily as a productivity multiplier over an existing skill base, while for someone still forming that base the same interaction doubles as a (potentially compromised) training signal. This is the reason this study isolates newcomers analytically rather than treating them as a subset of ``developers in general,'' as most prior work does.

\subsection{AI-Mediated Learning and Why Judgment Becomes Central}
\label{sec:background-learning}

The concept of \emph{computational thinking}, formulated by \citet{Win06}, organizes core software engineering skills around abstraction (separating the essential from the incidental) and decomposition (breaking problems into smaller, tractable parts). These skills are traditionally built in situations where the learner must engage with a task's difficulty: reading documentation, debugging, forming hypotheses, failing, and correcting course. Generative AI changes the paths through which a newcomer reaches an answer, and, in doing so, changes what gets practiced. \citet{Mah25} report that AI-assisted groups feel more confident and perceive tasks as less complex in the short term, with preliminary evidence of weaker unassisted performance afterward. \citet{Hib25} report that developers aged 16--25 estimate that 70--80\% of their programming activity is AI-supported, and describe, in their own words, effects on learning depth and command of fundamentals. The literature names this effect in different ways, shallow learning \citep{Mah25}, cognitive dependency \citep{Hib25}, retention loss \citep{Mah25}, skill erosion through over-reliance \citep{Szo25}, but converges on the same mechanism: the cognitive effort invested in searching for an answer is what used to sustain retention, and when AI removes that effort, learning tends to reorganize around new, faster, shallower criteria.

Confidence and autonomy sit in a related tension. \citet{Mah25} note that participants who feel more confident in the moment of AI use report lower confidence when later asked to apply the same skill unassisted, a tension they observe but do not name. \citet{Hib25} documents a parallel concern among young developers about the erosion of critical thinking through \emph{over-reliance}, particularly risky for beginners who may not yet have the knowledge needed to critically evaluate an AI answer \citep{Hib25}. Both threads motivate this study's attention, in RQ2 and RQ3, to how newcomers experience and manage these effects in their own words, rather than only measuring them experimentally.

\subsection{Market Reconfiguration and the Rise of Critical Judgment}
\label{sec:background-market}

AI assistants have also been described as reconfiguring the competencies expected of software developers. \citet{Doh23} characterize the phenomenon as a \emph{sea change} in the developer lifecycle, with effects on productivity, team shape, and hiring profiles; \citet{Szo25} describes an analogous shift while studying Copilot, framing it as a reorganization of valued competencies toward judgment, architecture, and technical communication, and away from raw code-writing. This has a direct implication for entry-level profiles: tasks historically associated with internship and junior roles, CRUD operations, scripts, small refactors, are among the tasks most frequently delegated to AI \citep{Szo25}. As a consequence, the entry barrier tends to rise, and the skill set expected of a newcomer shifts toward competencies previously associated with more senior levels, such as critical code review, architectural decisions, and translating business rules \citep{Doh23,Hib25}. The literature also points to a horizontal reorganization of the professional pyramid, with non-developer roles (product, data, business) gaining, through conversational interfaces, the ability to operate in what used to be exclusively technical territory \citep{Doh23,Szo25}. Together, these three threads, the shift in what AI is (Section~\ref{sec:background-ai}), the shift in how learning happens (Section~\ref{sec:background-learning}), and the shift in what the market rewards (this section), set up the central tension this study investigates: the same AI-mediated learning that trades depth for speed is unfolding at the exact moment the market is raising the value of the deep, judgment-based competencies that speed tends to erode. Section~\ref{sec:related} situates this study against five prior empirical studies that touch parts of this tension, and Section~\ref{sec:paradox} returns to it directly as the study's main theoretical contribution.

\subsection{Related Work}
\label{sec:related-work}

Five prior studies address AI assistant use by developers. \citet{Mah25} investigate, through a lab experiment still in progress, how ChatGPT use affects computational-thinking skills (abstraction and decomposition) in programming students, finding a confidence tension: AI-assisted participants feel more confident and perceive tasks as less complex, but report lower confidence in applying the skills later without assistance, alongside a frequent ``direct solution'' pattern associated with shallow learning. \citet{Hib25} conducted a qualitative study with developers aged 16--25, reporting that 70--80\% of participants' programming activities are AI-supported and discussing effects on learning depth and command of fundamentals, framed through the concept of \emph{over-reliance}. \citet{Kle24} studied professional developers' AI-assistant practices with a focus on security, documenting a systematic division of tasks driven by critical evaluation of what can be verified. \citet{Szo25} examined GitHub Copilot's impact on productivity, code quality, and ethics, organizing productivity into three dimensions (speed, quality, and perceived cognitive load) and describing a shift in valued competencies toward judgment, architecture, and technical communication. \citet{Pin24} studied developer experience with a contextualized AI coding assistant, describing task delegation patterns and the centrality of review to perceived quality.

Table~\ref{tab:related} compares these studies with the present work. Three gaps motivate this study: a scarcity of qualitative studies centered on interns and junior developers in active professional practice (existing studies use either controlled student samples \citep{Mah25} or already-established professionals \citep{Kle24,Pin24}); the absence of an integrated reading that connects usage patterns, learning, autonomy, and market perception within a single analysis; and a lack of a rich description of the individual self-regulation strategies that newcomers report in real, everyday use.

\begin{table}[h]
	\caption{Comparison between this study and five related studies.}
	\label{tab:related}
	\centering
	\small
	\begin{tabular}{p{2.6cm}p{2.6cm}p{2.6cm}p{5.5cm}}
		\toprule
		\textbf{Study} & \textbf{Population} & \textbf{Method} & \textbf{Main focus} \\
		\midrule
		\citet{Mah25} & Undergraduate students & Lab experiment & Computational thinking and prompting patterns \\
		\citet{Hib25} & Developers, 16--25 y.o. & Qualitative interviews & Technical training and over-reliance \\
		\citet{Kle24} & Professional developers & Qualitative interviews & Usage practices and security \\
		\citet{Szo25} & Developers in general & Review and qualitative analysis & Productivity, quality, and ethics with Copilot \\
		\citet{Pin24} & Developers in general & Qualitative experience study & Usability and expectations with contextualized AI \\
		\textbf{This study} & \textbf{Interns and juniors, in practice} & \textbf{Qualitative, thematic analysis \citep{BC06}} & \textbf{Usage patterns, learning, autonomy, and market, from newcomers' own perspective} \\
		\bottomrule
	\end{tabular}
\end{table}

\section{Method}
\label{sec:method}

\subsection{Research Design}
This is a qualitative, descriptive, and exploratory study, following an inductive approach. Data were collected through semi-structured interviews and analyzed with the six-phase Thematic Analysis method proposed by \citet{BC06}. Thematic analysis was preferred over alternative qualitative approaches (e.g., grounded theory or content analysis) for two reasons specific to this study's aims. First, it does not require, and in fact discourages, committing to an \emph{a priori} theoretical framework before coding, which suits an exploratory question with no established model of how AI reshapes early-career software work. Second, it is explicitly designed to surface patterns of shared meaning across a moderate-sized interview set without requiring the kind of large, continuously-sampled corpus that grounded theory's saturation criteria typically demand, which matches the scope of an undergraduate research project constrained in time and access to participants.

\subsection{Participants}
Participants were selected by intentional and convenience sampling, contacted through the researcher's personal network and, complementarily, through LinkedIn. The target profile comprised interns and junior developers working at software companies in Brazil. No minimum professional tenure was required, and AI tool use was not an explicit inclusion criterion, although all thirteen participants who took part turned out to use such tools in their daily work. The study reached thirteen interviews. This number was not fixed in advance; interviewing continued until new transcripts stopped introducing codes not already present in the codebook and instead repeated patterns already seen in prior interviews, a practical proxy for thematic saturation given the project's resource constraints. By the last two or three interviews, coding was producing near-exclusively re-applications of existing codes rather than new ones, which was taken as the stopping signal; this also means the sample size reflects saturation for the depth of analysis pursued here, not a target number chosen up front. Nine interviews were conducted by the first author; the remaining four were conducted by a second interviewer working on a related master's research topic, with participants' explicit authorization for data use in this study. Table~\ref{tab:participants} summarizes participant profiles; names were replaced with codes E01--E13 to preserve anonymity, and companies are described only by industry sector.

\begin{table}[h]
	\caption{Participant profile.}
	\label{tab:participants}
	\centering
	\small
	\begin{tabular}{lp{3.6cm}p{2.4cm}p{3.6cm}c}
		\toprule
		\textbf{Code} & \textbf{Role} & \textbf{Tenure} & \textbf{Company sector} & \textbf{Gender} \\
		\midrule
		E01 & Junior developer (back-end) & $\sim$2 years & Software house & M \\
		E02 & Junior developer (back-end) & 8 months as junior & E-commerce & M \\
		E03 & Intern (DevOps/cloud) & $\sim$1.5 years & Technology & M \\
		E04 & Intern (full-stack) & $\sim$2 years & R\&D technology institute & M \\
		E05 & Junior developer & 1.5 years & Technology & M \\
		E06 & Junior developer (back/front) & $\sim$2 years & Software house (banking) & M \\
		E07 & Junior developer (full-stack) & $\sim$6 months as junior & Fintech & M \\
		E08 & Junior developer (back-end) & 2 years & IT consulting & M \\
		E09 & Junior developer (data/support) & 1.5 years & Consulting (financial sector) & M \\
		E10 & Junior software engineer & 1 year & Large technology company & F \\
		E11 & Intern & $\sim$8 months & Technology & M \\
		E12 & Intern (full-stack web) & $\sim$4 months & Software house & F \\
		E13 & Intern & $\sim$4 months & Software house & M \\
		\bottomrule
	\end{tabular}
\end{table}

\subsection{Data Collection}
Semi-structured interviews were conducted individually via videoconference (Google Meet, Microsoft Teams, or Zoom) and fully transcribed. The interview guide was organized into thematic blocks aligned with the four sub-questions, favoring open questions that encouraged participants to report concrete situations from their daily work. Verbal consent for recording and for research use was obtained at the start of each interview, after the researcher explained the study's purpose, procedure, and confidentiality policy.

\subsection{Thematic Analysis Procedure}
Analysis followed the six phases proposed by \citet{BC06}: (1) familiarization with the data through full transcription and repeated reading; (2) inductive generation of initial codes; (3) search for themes, grouping codes into candidate themes with the aid of concept maps; (4) review of themes at both the coded-extract and full-dataset levels; (5) definition and naming of final themes; and (6) report production, presented in Section~\ref{sec:results}.

Coding stability across interviewers was checked, rather than assumed, by design. The codebook was built inductively from the first author's nine interviews only. It was then applied, as-is, to the four interviews conducted by the second interviewer, with an explicit rule that no new code could be created during this second pass unless the existing codebook genuinely failed to capture a segment of data. In practice, the four additional interviews were fully coded using only pre-existing codes, with no new codes required, which is the concrete evidence behind the claim of coding stability: it means the thematic structure built from one interviewer's data generalized, without modification, to data collected independently by a second interviewer on a different, if related, research question. This procedure is a lightweight analogue to inter-coder reliability checks used in larger qualitative studies, adapted to the resources available for an undergraduate thesis.

\subsection{Ethical Considerations}
The study followed standard ethical principles for research with human subjects. Participation was voluntary, without financial compensation, and no sensitive personal data (as defined by Brazil's General Data Protection Law) were collected. Participant identity was preserved through the use of codes (E01--E13) in analysis and reporting, and details that could enable direct identification (proper names, company, project, and product names in sensitive contexts) were suppressed or replaced in transcripts and quotations.

\section{Results}
\label{sec:results}

This section presents the thematic analysis of the thirteen interviews, organized by research question, with illustrative quotations and divergent cases. Participants are identified by codes E01--E13. Section~\ref{sec:synthesis} integrates the four sub-answers into a single reading.

\subsection{RQ1: Usage Patterns and Decision Criteria}
\label{sec:rq1}

\paragraph{Verification-conditioned use.} Across the thirteen interviews, the main criterion participants use to decide between using and not using AI was neither deadline nor task complexity, but the possibility of checking the result. They turn to AI for what they know how to review, and avoid it where they cannot judge whether the answer is correct. E10 states the criterion at both ends: \enquote{I only use AI for things I already know how to do, because then I'm able to judge the result. I'll never use it for something I've never done in my life, because then it will produce whatever, and I won't know if it's right or wrong} (E10). E06 names the same rule as the explicit decision factor: \enquote{what makes me decide to use it or not is how much command I have over what I need to do} (E06). E01 anchors it in a concrete example: \enquote{I don't trust anything I can't validate [...]. If I don't know how to work with Terraform, don't know how to configure AWS, I'm not going to trust it blindly} (E01).

This criterion inverts a common expectation: it is not the newcomer with the smallest repertoire who relies most on AI to cover knowledge gaps. In the reports, the opposite happens, since without the repertoire to judge the result, the task falls on the side where AI is avoided. Two cases test the rule: E11 used AI to integrate the Google Maps API without mastering the topic, asked a colleague to review it, and still felt uneasy; E07 let AI implement a business rule in a financial project without mastering the rule, the client noticed the error, and the team had to roll back operations. These cases do not weaken the criterion; they show what happens when it is ignored.

\paragraph{What is delegated to AI and what is kept.} A recurring split separates tasks where AI is used from tasks where it is avoided (Table~\ref{tab:tasks}). The left column groups tasks with a known right answer or standard pattern; the right column groups tasks that depend on project context and history. E02 frames the split: \enquote{I prefer to avoid it for tasks related to business rules, because I want to stay familiar with what the project is doing} (E02). E07 evades from the opposite end of the same rule, familiarity: \enquote{if it's a library I already know, I won't hand it to AI}, adding that \enquote{when the system is multi-repo, AI fails, it doesn't have the full context} (E07). The one participant who departs from this map, E13, simply does not use AI at work, justifying the refusal because \enquote{AI has a way of writing that you can identify} (E13); even so, the same typology appears in negative form.

\begin{table}[h]
	\caption{Division of tasks in AI use, as reported by the 13 participants.}
	\label{tab:tasks}
	\centering
	\small
	\begin{tabular}{p{6cm}p{6cm}}
		\toprule
		\textbf{Tasks where AI is used} & \textbf{Tasks where AI is avoided} \\
		\midrule
		Front-end, HTML, layout & Architecture decisions \\
		CRUD and simple data operations & Business rules \\
		Recalling one-off syntax & Broad, multi-repository context \\
		Generating automated tests & Infrastructure configuration and logging \\
		Initial project scaffolding & Tasks that are faster to do by hand \\
		External libraries not worth studying & \\
		Debugging and reading existing code & \\
		Refactoring and code summarization & \\
		\bottomrule
	\end{tabular}
\end{table}

\paragraph{AI as an accessible senior.} In every interview that touched this point, AI appears as a partial substitute for the senior colleague, covering two roles: explaining concepts on demand and helping to think before implementing. E08 states this preference directly: \enquote{in most cases, I'd rather ask artificial intelligence for help than ask someone} (E08). The most consistent effect, however, was not reducing conversation with seniors but changing the type of question taken to them. E01 describes the shift: \enquote{the question goes from `how do I implement this?' to `is what I implemented right according to the company's rule?'} (E01). E03 confirms from another angle: \enquote{the questions I used to bring before AI were more about code. The questions I bring today are usually about design, about architecture} (E03). The divergent case is E08, who reports asking seniors the same kinds of questions as before, showing that the shift depends on how much each professional integrates AI into their routine; it is not automatic.

\paragraph{Company context shapes how the criterion is applied.} The criteria above are not purely individual; participants fall into four positions regarding organizational context (Table~\ref{tab:company}). E10 describes the pro-AI pole, with structured training and paid tools: \enquote{it's not `AI First' in the sense that AI will do everything, it's the opposite: let's use it to improve our process [...] the more you study how the tool works, the less faith you have in it} (E10). E09 describes the opposite pole: \enquote{my team is led by an older gentleman. He doesn't really like the idea of using AI. He taught me by hand} (E09). Both cases operate the same underlying criterion (using only what one can validate); what changes is where the confidence and vocabulary come from.

\begin{table}[h]
	\caption{Participants' position regarding the organizational context of AI use.}
	\label{tab:company}
	\centering
	\small
	\begin{tabular}{p{3.6cm}p{2.4cm}p{6.4cm}}
		\toprule
		\textbf{Position} & \textbf{Participants} & \textbf{Characteristic} \\
		\midrule
		Pro-AI company with structured training & E10 & Training tracks, external specialists, company-paid tools \\
		Productivity pressure via AI & E05, E06, E07 & Deadlines and story points already assume AI use \\
		Allowed but unstructured use & E01, E02, E03, E04, E08, E11, E12, E13 & Company allows it, without training or requiring it \\
		Leadership against AI & E09 & Senior lead taught the participant ``by hand,'' avoiding AI \\
		\bottomrule
	\end{tabular}
\end{table}

\paragraph{RQ1 summary.} The decision criterion between AI and manual work is not deadline or complexity, but the ability to check the result. This criterion separates tasks (CRUD, front-end, syntax, and testing on one side; architecture, business rules, and broad context on the other) and shifts the type of question newcomers bring to seniors.

\subsection{RQ2: Reconfiguration of the Learning Process}
\label{sec:rq2}

\paragraph{AI changes the path to knowledge.} Before AI, technical doubts were resolved through documentation, forums (Stack Overflow), and more experienced colleagues. Participants report that these paths did not disappear, but lost priority: AI became the first resource consulted, since it answers each participant's specific doubt directly. E02 explains why: \enquote{you ask something and it answers, and you just do it. It's much more convenient than reading documentation, searching for where you're going wrong} (E02). E01 translates the day-to-day effect: \enquote{I practically don't look at documentation anymore, unless it's something extremely specific} (E01). E12 calls this new route \enquote{a tutor that's there for you, and that won't give you the same answer it would give everyone else} (E12).

\paragraph{Shallower learning, with individual defenses.} Most participants say that AI-mediated learning tends to be shallower than learning by reading, research, and trial and error; depth is traded for speed. E02 states this most sharply: \enquote{I feel like I'll never be as experienced, never be as good at programming as someone who grew up without artificial intelligence. What it teaches is much more forgettable than what I'd learn if I made a mistake and had to go to Stack Overflow to figure it out} (E02). E04 gives the underlying mechanism: \enquote{when you ask it to do something and you just review it, I believe that ends up creating a dependency} (E04). E10 compares it to school: \enquote{it would be like when we used to copy a school assignment from Wikipedia} (E10). The defense against this risk is individual: some adopt clear rules, e.g., E10 (\enquote{I don't use any function I haven't understood the reason for}) and E13 (\enquote{a big part of how I learn is understanding why something works}); this division does not track career length, since E10 has one year as a junior and E13 has four months, yet both apply the same defense.

\paragraph{Loss of cognitive friction.} Several participants describe, in different words, the same phenomenon: the mental effort once needed to find an answer is no longer exerted, and that effort was, according to them, part of what sustained learning. E04 articulates the mechanism: \enquote{the faster the answer comes, the easier it gets, the less our mind needs to prioritize storing that information} (E04). E10 extends this to the long term: \enquote{I think it takes away some of people's resilience. Because when you don't use AI, you go read the article, read the code snippet, read the documentation, keep debugging to see the behavior. That makes you more resilient, because you know not everything gets solved so fast} (E10). E11 names another facet of the loss: \enquote{sometimes it feels like, without AI, you wouldn't be doing so well. There's a sense that the credit belongs more to the tool than to the person} (E11).

\paragraph{Questions brought to seniors have changed.} In 12 of the 13 reports, questions about how to implement something, which syntax to use, or which library to choose disappear, since AI already answers them; what remains for seniors are questions of project convention, business rules, and architectural decisions. E07 confirms: \enquote{it ends up being much more about business rules. Technical doubts I used to bring, like modeling questions, I don't bring anymore} (E07). E04 notes a secondary effect: \enquote{you arrive with a much more elaborate question. Instead of saying `I can't get this to work', I'll say `I can't get this to work because of this and that'} (E04).

\paragraph{RQ2 summary.} AI reconfigures learning in three ways: it replaces Google, Stack Overflow, and documentation as the initial route; it reduces the depth of knowledge built, defended against mainly through individual posture; and it removes the cognitive effort that sustained learning, with consequences for resilience and perceived authorship. The same shift appears in questions brought to seniors, moving from ``how do I do this'' to ``is this good enough?''

\subsection{RQ3: Autonomy, Trust, and Perceived Dependence}
\label{sec:rq3}

\paragraph{The autonomy paradox.} Most participants describe AI as a source of autonomy: they can read code they did not know, finish tasks in languages they do not yet master, integrate libraries they never used. At the same time, several recognize that this autonomy is partly borrowed from the tool. E03 gives the starkest version of the paradox: \enquote{with it, I'm very autonomous, I swear, I'm capable of anything, I could probably send a rocket to NASA. But without it, no} (E03). E02 reports the opposite feeling, less autonomy: \enquote{I think I'm actually less autonomous. Because it's so easy to grab information, solve your code, deliver things that are good, fast, well-written and readable, I'm never going to be as competent a programmer as people who were trained without artificial intelligence} (E02). E11 names the displaced authorship: \enquote{that would be attributing to myself credit that belongs to the tool. There isn't really a creation of my own there} (E11).

\paragraph{Trust comes from review, not from the output.} Trust in the AI's result does not arrive when the answer is generated; it arrives afterward, once the participant reviews it. E08 gives the simplest version: \enquote{if it's a simple task, just a sum in the middle of it, I let it pass, I trust it, although I test it later. If it was highly complex, I stop to do some review and test it several times} (E08). E10 describes review as a dialogue with the tool itself: \enquote{it's kind of like a virtual rubber duck. You go along talking to it, it's kind of like talking to yourself. You ask, you look, then you think, and you're able to critique it better} (E10). Trust operates conditionally: it depends on reading, comparison with existing knowledge, and testing; when this step is skipped, participants report discomfort, as in E11's Google Maps case (Section~\ref{sec:rq1}).

\paragraph{Dependence is denied in the first person.} When asked whether AI creates dependence, nearly all participants said yes in general, and nearly all denied it happens to them. E09 illustrates this within a single interview: about others, AI \enquote{takes away critical thinking} and colleagues become \enquote{hostage to a prompt} (E09); about himself: \enquote{currently, no [I don't have dependence]. I use it to explain and to learn. A tool that empowers, but shouldn't be allowed to take over} (E09). Two cases break the pattern: E02 admits fear directly (\enquote{I'm afraid that, in more complex situations, I'll feel lost}), and E03 admits dependence openly in the first person (\enquote{just thinking, `I'll stop using Claude, I'll have to use just GPT', man, that alone would wreck me}) (E03).

\paragraph{Self-regulation strategies.} The defense against the risks above operates individually, through four recurring strategies: (1) reviewing all output before accepting it (E04: \enquote{I like to invest a bit more time in the review part, mainly understanding what's being done, what approach it took and why}); (2) refusing to use AI without understanding it (E13: \enquote{a big part of how I learn is understanding why something works}); (3) asking the AI itself for explanations rather than accepting the first answer (E11: \enquote{I ask a lot and even argue against its suggestions. If I don't understand something, I question whether that solution wouldn't make the process slower, or if it's redundant}); and (4) keeping parallel practice, such as individual study, exercises on platforms like LeetCode, or deliberately choosing to do some tasks by hand (E09: \enquote{I take the code, study it, and change mine, so I don't just take all the changes}).

\paragraph{RQ3 summary.} AI increases perceived autonomy for already-mastered tasks, but trust comes from review rather than from the output itself, and dependence is denied in the first person even when acknowledged in general. Self-regulation, reviewing before accepting, refusing to use AI without understanding it, asking the tool for explanations, and maintaining parallel practice, is what sustains this arrangement.

\subsection{RQ4: Reconfiguration of Market Entry}
\label{sec:rq4}

\paragraph{The entry barrier has risen.} There is near-universal consensus that entering the market has become harder for interns and juniors in recent years. E04 captures the tension almost ironically: \enquote{you're already expected to be an intern with years of experience, already expected to be a junior with that kind of experience. So how are you supposed to get experience if you haven't had experience yet?} (E04). E07 names the direct cause: \enquote{the entry barrier increased a lot, because companies demand more from you, whether they mean to or not, because AI builds the code that juniors used to write} (E07). E10 explains from the company's side: \enquote{the market trend is to reduce the number of devs a bit, have more automation. But the devs who stay will be the ones who know how to work with AI} (E10). This finding deserves an attribution caveat: reports capture newcomers' perception of the barrier and of the cause they attribute to it, not an isolated measure of AI's effect on the market. During the period studied, the rise in the perceived barrier coincides with a broader contraction in hiring for entry-level technology roles, associated with macroeconomic factors and layoff cycles that are prior to, and independent of, generative AI (Section~\ref{sec:threats}). Not every account attributes the difficulty to AI: E06, E11, and E12 describe a generally tight market, while explicit attribution to AI appears in E07 and E10.

\paragraph{The differentiator shifted from writing to evaluating code.} If writing code is no longer a differentiator, participants converge on what replaced it: the ability to judge code, that is, to recognize when AI output is wrong, ignores a business rule, or breaks a project's conventions. E07 puts it plainly: \enquote{the differentiator, in my view, is knowing technologies that AI doesn't handle so well} (E07). E11 describes the shift from the job-market side: \enquote{it used to be that knowing how things worked and writing the code by hand was the big differentiator. Today, knowing how to write code has become commonplace} (E11), adding that \enquote{the focus seems to be much more on how you reason, on concepts, and on knowing what should be present in the solution} (E11). E05 summarizes: \enquote{it's increasingly going to be about the business, and less about code. You'll need to understand the business much more, the architecture, how to organize the project. How to do it, not necessarily doing it} (E05). This value shift (from writing to judging) creates a delicate problem for newcomers: they must now exercise a competence (critical judgment) that used to be built precisely by writing code, a task they now partly hand over to AI. This \emph{formative paradox} is discussed in Section~\ref{sec:paradox}.

\paragraph{Reorganization of the professional pyramid.} Reports describe three shifts across the career ladder. First, \emph{seniors become orchestrators}: E07 notes that \enquote{if a senior knows how to operate AI well, they'll produce the code a junior would make. A single senior can sometimes manage an entire team} (E07); E10 describes this practice directly: \enquote{I can put together a squad of agents here. I've got five, six tasks to do, I let it work, then I come back reviewing, checking, testing. Since I already have the ability to judge the code, that makes it easier for me} (E10). Second, \emph{AI brings non-developers into technical territory}: E10 describes a Product Manager performing operations that used to depend on a developer, via a conversational agent that queries an API on her behalf. Third, \emph{juniors face internal competition}: companies tend to prefer a senior using AI over several juniors, since the cost of training a junior becomes large relative to the productivity of a well-equipped senior.

\paragraph{Three coexisting readings of the future.} Despite the tensions above, participants' emotional reading of the future is not uniform. An \emph{optimistic} stance sees AI as a tool that will make work more strategic (E04: \enquote{the dev's work will become more strategic, more about thinking than doing}; E09: it will \enquote{give us superpowers to work better}). A \emph{concerned} stance foresees a harder path, especially at entry level (E02: \enquote{if we outsource understanding of the business to AI, when something goes wrong with a business rule we wrote, we won't know how to fix it or even explain to the AI what's happening}; E12: \enquote{they're asking for so much, so much. I don't have all of that}). A \emph{pragmatic} stance accepts the change with caution; E13 offers the most vivid metaphor: \enquote{it's like the invention of the car: there's no taking the car out of circulation. But it's like this: use it with caution. A car can be a weapon, but it can also be a means of transportation. AI is the same thing} (E13).

\paragraph{RQ4 summary.} The entry barrier has risen, and participants associate part of this rise with the migration of entry-level tasks to AI, without ruling out a broader labor-market contraction as a contributing factor. The differentiator moved from writing to evaluating code, the professional pyramid is reorganizing (seniors orchestrating agents, non-developers entering technical territory), and three postures toward the future coexist: optimistic, concerned, and pragmatic.

\subsection{Integrated Synthesis}
\label{sec:synthesis}

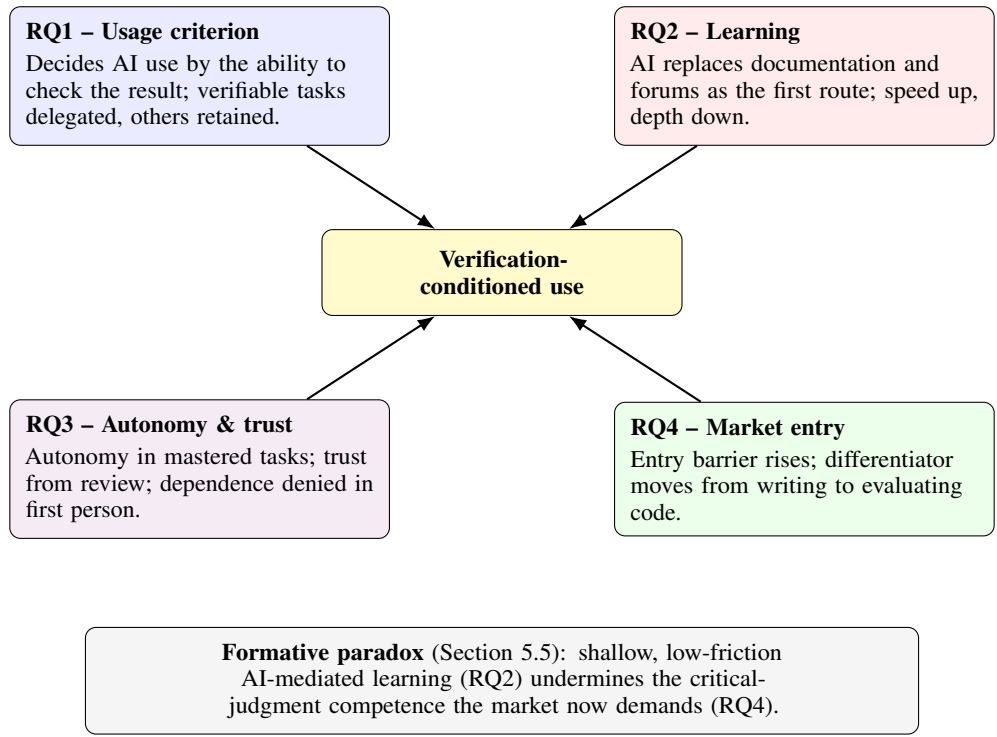
\begin{figure}[h]
	\centering
	\begin{tikzpicture}[
		box/.style={draw, rounded corners, align=left, text width=4.6cm, font=\small, inner sep=6pt},
		center_box/.style={draw, rounded corners, align=center, text width=4.2cm, font=\small\bfseries, fill=yellow!25, inner sep=8pt},
		arr/.style={-{Latex}, thick}
		]
		\node[box, fill=blue!8] (rq1) at (0,5) {\textbf{RQ1 -- Usage criterion}\\[2pt] \normalfont Decides AI use by the ability to check the result; verifiable tasks delegated, others retained.};
		\node[box, fill=red!8] (rq2) at (8,5) {\textbf{RQ2 -- Learning}\\[2pt] \normalfont AI replaces documentation and forums as the first route; speed up, depth down.};
		\node[center_box] (center) at (4,2.4) {Verification-\\conditioned use};
		\node[box, fill=violet!8] (rq3) at (0,-0.2) {\textbf{RQ3 -- Autonomy \& trust}\\[2pt] \normalfont Autonomy in mastered tasks; trust from review; dependence denied in first person.};
		\node[box, fill=green!8] (rq4) at (8,-0.2) {\textbf{RQ4 -- Market entry}\\[2pt] \normalfont Entry barrier rises; differentiator moves from writing to evaluating code.};
		\draw[arr] (rq1) -- (center);
		\draw[arr] (rq2) -- (center);
		\draw[arr] (rq3) -- (center);
		\draw[arr] (rq4) -- (center);
		\node[box, fill=gray!8, text width=10.6cm, align=center] (paradox) at (4,-3) {\textbf{Formative paradox} (Section~\ref{sec:paradox}): shallow, low-friction AI-mediated learning (RQ2) undermines the critical-judgment competence the market now demands (RQ4).};
	\end{tikzpicture}
	\caption{Integrated synthesis of the results: convergence of the four research questions around verification-conditioned use.}
	\label{fig:synthesis}
\end{figure}

Table~\ref{tab:synthesis} condenses the four sub-answers. Results converge on \textbf{verification-conditioned use} (Section~\ref{sec:rq1}) as a central idea that reappears across all four research questions: it is the usage criterion in RQ1, it explains who learns with depth in RQ2, it is sustained by the self-regulation strategies in RQ3, and it becomes the market's valued differentiator in RQ4 (Figure~\ref{fig:synthesis}).

\begin{table}[h]
	\caption{Synthesis of answers to the four research questions.}
	\label{tab:synthesis}
	\centering
	\small
	\begin{tabular}{lp{2.6cm}p{8.5cm}}
		\toprule
		\textbf{RQ} & \textbf{Focus} & \textbf{Key finding} \\
		\midrule
		RQ1 & Usage criterion & AI use is decided by the ability to check the result; this separates tasks, delegating what is verifiable and retaining what cannot be judged. \\
		RQ2 & Learning & AI becomes the initial route instead of Google, Stack Overflow, and documentation; speed is gained and depth and cognitive effort are lost; questions to seniors move from ``how do I do this'' to ``is this correct?'' \\
		RQ3 & Autonomy, trust, dependence & More autonomy in mastered tasks; trust comes from review, not from output; dependence is admitted in general and denied in the first person; self-regulation sustains the arrangement. \\
		RQ4 & Market entry & The entry barrier rises as typical junior tasks become AI's work; the differentiator shifts from writing to evaluating code; optimistic, concerned, and pragmatic postures coexist. \\
		\bottomrule
	\end{tabular}
\end{table}

\section{Discussion}
\label{sec:discussion}

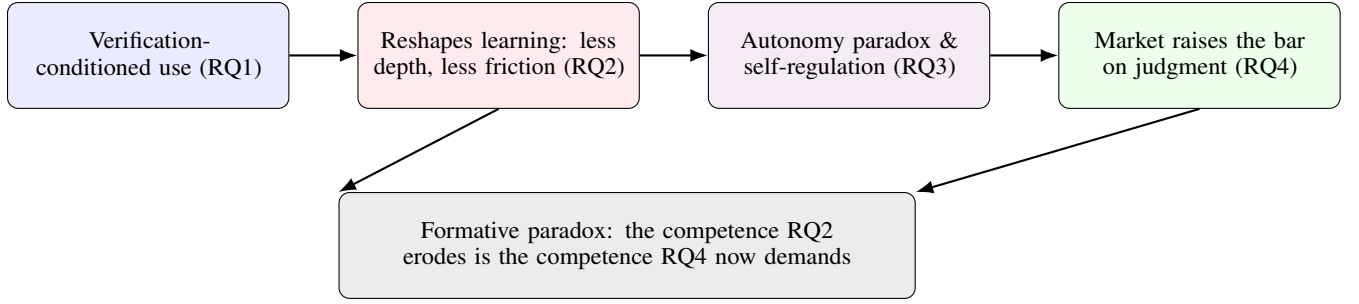
\begin{figure}[h]
	\centering
	\begin{tikzpicture}[
		node distance=0.9cm,
		chainbox/.style={draw, rounded corners, align=center, text width=3.3cm, font=\small, inner sep=6pt, minimum height=1.4cm},
		arr/.style={-{Latex}, thick}
		]
		\node[chainbox, fill=blue!8] (s1) {Verification-conditioned use (RQ1)};
		\node[chainbox, fill=red!8, right=of s1] (s2) {Reshapes learning: less depth, less friction (RQ2)};
		\node[chainbox, fill=violet!8, right=of s2] (s3) {Autonomy paradox \& self-regulation (RQ3)};
		\node[chainbox, fill=green!8, right=of s3] (s4) {Market raises the bar on judgment (RQ4)};
		\node[chainbox, fill=gray!15, below=1.1cm of s2, xshift=1.7cm, text width=7.2cm] (s5) {Formative paradox: the competence RQ2 erodes is the competence RQ4 now demands};
		\draw[arr] (s1) -- (s2);
		\draw[arr] (s2) -- (s3);
		\draw[arr] (s3) -- (s4);
		\draw[arr] (s2.south) -- (s5.north west);
		\draw[arr] (s4.south) -- (s5.north east);
	\end{tikzpicture}
	\caption{Logical chain from the usage criterion to the study's theoretical contribution.}
	\label{fig:chain}
\end{figure}

This section discusses the findings of Section~\ref{sec:results} against the literature reviewed in Section~\ref{sec:related}, following the same four research questions, and closes with the \textbf{formative paradox}, the study's main theoretical contribution. Figure~\ref{fig:chain} previews the argument as a single causal chain: the verification criterion identified in RQ1 is what reshapes learning in RQ2; that reshaped learning is what produces the autonomy paradox and the self-regulation response in RQ3; and, in parallel, the market shifts documented in RQ4 raise exactly the competence that RQ2's shallower learning puts at risk, which is what the formative paradox names.

\subsection{RQ1: Usage Patterns and the Verification Criterion}
\textbf{Confirms.} The central RQ1 finding, that the decision between AI and manual work hinges on the ability to check the result, converges with \citet{Kle24}, who report a systematic division of tasks among professional developers driven by critical evaluation of what can be verified. It also converges with \citet{Pin24} on the importance of review for perceived quality. \textbf{Contradicts.} There is tension with \citet{Hib25} and \citet{Mah25}, who tend to treat newcomers' limited repertoire as a vulnerability leading them to accept wrong answers; the present reports show the opposite, newcomers exercise more caution precisely because the cost of error is higher for them, illustrated by E07's rolled-back business-rule error. \textbf{Extends.} This study extends \citet{Kle24}'s pattern, observed in professional developers, to the very start of the career, showing it is not inherited from seniority but built independently by each newcomer; and it extends \citet{Pin24} by showing that review is not just quality control but the condition that makes AI use sustainable at all (Section~\ref{sec:paradox}).

\subsection{RQ2: Reconfiguration of Learning}
\textbf{Confirms.} The dialogue with \citet{Mah25} is most direct: their preliminary lab results point to lower perceived complexity and possibly worse later unassisted performance in AI-assisted groups, echoed here by participants naming the phenomenon in the first person (E02). There is also convergence with \citet{Hib25} on \emph{over-reliance} as an explanatory category. \textbf{Contradicts.} These reports diverge from \citet{Hib25} on what sustains the defense against over-reliance: Hibi tends to treat the problem as generational, while self-regulation here is individual and does not track career length (E13, four months, applies the same strategies as E10, one year). \textbf{Extends.} The contrast with \citet{Win06}'s classical formulation of computational thinking, in which abstraction and decomposition are built when the learner must handle a task's difficulty, is instructive: these reports show that AI is displacing exactly those difficulty-laden situations, an effect Wing's framework anticipates but does not itself describe for an AI-mediated context, and the participants counter it, uniquely, by deliberately maintaining parallel unassisted practice, a coping strategy absent from the cited literature.

\subsection{RQ3: Autonomy, Trust, and Dependence}
\textbf{Confirms.} The autonomy paradox (Section~\ref{sec:rq3}) is the qualitative counterpart of a tension \citet{Mah25} observe at a different level: their participants feel more confident in the moment and less confident later, while these participants feel more capable and less in ownership of the result at the same time. \textbf{Contradicts.} On dependence, this study diverges most from \citet{Hib25}: while over-reliance is generally treated as an observable, self-admitted phenomenon, these reports show a consistent pattern of recognizing dependence in others while denying it in oneself. \textbf{Extends.} This repositions \citet{Mah25}'s account of trust as a moment-of-generation state, showing instead that trust is built afterward, through reading, comparison, and testing; and it extends \citet{Hib25}'s over-reliance construct by suggesting dependence is treated by newcomers as a collective risk to be monitored, not a stable individual trait, a distinction the source literature does not draw.

\subsection{RQ4: Reconfiguration of the Market}
\textbf{Confirms.} The shift of the professional differentiator from writing to evaluating code is compatible with \citet{Szo25}'s account of competencies reorganizing toward judgment, architecture, and technical communication, and the reorganization of the professional pyramid (seniors orchestrating agents) is compatible with \citet{Szo25}'s productivity findings for experienced developers. \textbf{Contradicts.} The rise in the entry barrier is in tension with \citet{Doh23}, whose GitHub-affiliated report projects that AI democratizes entry and benefits less experienced developers most, the opposite of what interviewees perceive; it aligns instead with \citet{Szo25}'s prediction that automation may reduce demand for entry-level positions. \textbf{Extends.} This study's contribution is showing that the competency reorganization described by \citet{Szo25} at the market level is perceived by newcomers themselves, not only inferred by employers or analysts; and the reorganization it documents, non-developers entering technical territory via conversational interfaces, is a movement not fully described in the literature reviewed here, opening a line for future investigation.

\subsection{The Formative Paradox}
\label{sec:paradox}

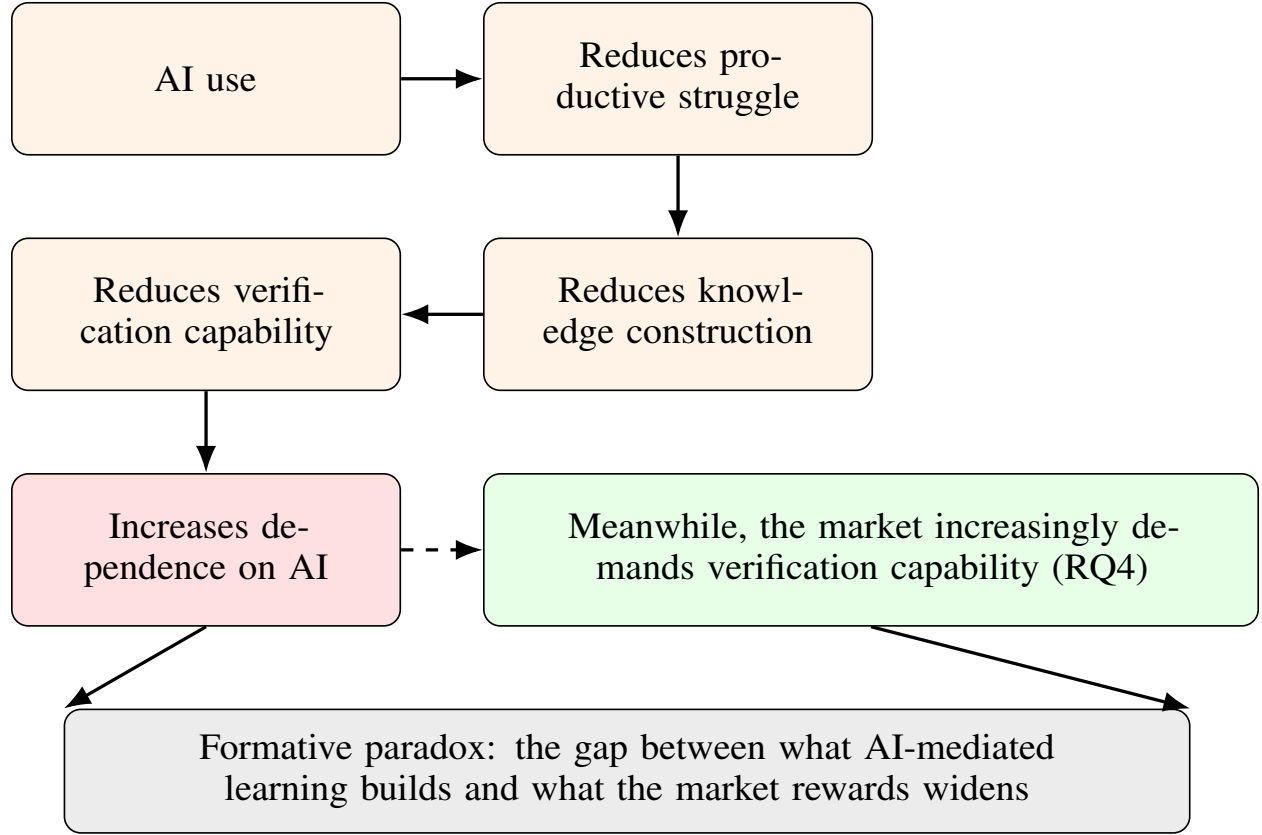
\begin{figure}[h]
	\centering
	\resizebox{\textwidth}{!}{%
	\begin{tikzpicture}[
		node distance=0.7cm and 0.7cm,
		mstep/.style={draw, rounded corners, align=center, text width=2.9cm, font=\small, inner sep=6pt, minimum height=1.3cm},
		arr/.style={-{Latex}, thick}
		]
		\node[mstep, fill=orange!10] (a1) {AI use};
		\node[mstep, fill=orange!10, right=of a1] (a2) {Reduces productive struggle};
		\node[mstep, fill=orange!10, below=of a2] (a3) {Reduces knowledge construction};
		\node[mstep, fill=orange!10, left=of a3] (a4) {Reduces verification capability};
		\node[mstep, fill=red!12, below=of a4] (a5) {Increases dependence on AI};
		\node[mstep, fill=green!10, right=of a5, text width=6.2cm, minimum height=1.3cm] (b1) {Meanwhile, the market increasingly demands verification capability (RQ4)};
		\node[mstep, fill=gray!15, below=of a5, xshift=3.6cm, text width=9.2cm, minimum height=1.0cm] (b2) {Formative paradox: the gap between what AI-mediated learning builds and what the market rewards widens};
		\draw[arr] (a1) -- (a2);
		\draw[arr] (a2) -- (a3);
		\draw[arr] (a3) -- (a4);
		\draw[arr] (a4) -- (a5);
		\draw[arr, dashed] (a5) -- (b1);
		\draw[arr] (a5.south) -- (b2.north west);
		\draw[arr] (b1.south) -- (b2.north east);
	\end{tikzpicture}%
	}
	\caption{Mechanism of the formative paradox: a self-reinforcing loop on the AI-use side runs against rising market demand for the very competence it erodes.}
	\label{fig:mechanism}
\end{figure}

Reading the four research questions together reveals a tension that organizes much of this study's findings: the \textbf{formative paradox}. In short: \emph{the same effects of AI-mediated learning, shallow and low in cognitive friction, make it harder to build the critical-judgment competence that the market has begun to demand as a professional differentiator.} This is not a logical contradiction but a counterintuitive situation in which two expected movements pull in opposite directions, in the same sense as the confidence tension observed by \citet{Mah25} and the autonomy paradox of Section~\ref{sec:rq3}. Figure~\ref{fig:mechanism} lays out the mechanism as a small causal model rather than a single finding: it is the specific chain from reduced productive struggle to reduced verification capability, running in parallel with a market that is raising, not lowering, its demand for that same capability, that makes the tension self-reinforcing rather than a one-off side effect.

The tension emerges when two separately unremarkable findings are read together. RQ2 shows that newcomers learn more shallowly, with less cognitive effort, losing the friction that historically sustained knowledge retention. RQ4 shows that the market has begun to value exactly the competence built through that friction: critical judgment, reading others' code, architectural decision-making. The consequence is direct: newcomers who accept AI's speed without conditioning its use on verification gain short-term productivity but lose the chance to build the competence the market will require of them in the medium term; newcomers who condition use on verification preserve that competence but pay in speed. This choice is not entirely free, it depends on company context, leadership, deadline pressure, and individual posture toward one's own training, which is why self-regulation (Section~\ref{sec:rq3}) is not a side trait but the piece that makes AI use sustainable from the newcomer's own point of view.

This tension engages with \emph{over-reliance} \citep{Hib25,Mah25} but is not exhausted by it: those accounts treat the effect as individual and cognitive, while this study's contribution is showing that the tension also sits between the newcomer and the market, which defines what counts as a differentiator. \citet{Doh23}'s account of a regime change in the developer lifecycle touches this point but does not descend to the level of individual training. Notably, the individual defenses \citet{Hib25} and \citet{Mah25} recommend in general terms, careful validation and staying actively engaged with the problem, are exactly what these newcomers report adopting on their own (E10, E01, E13), at a cost: the speed gained by passive acceptance is precisely what would help clear the entry barrier they must cross. This study does not resolve the paradox; it names it and describes how it appears, grounded in first-person accounts.

\subsection{Theoretical Contribution}
\label{sec:theory}

The discussion above yields a small conceptual model rather than a single isolated finding, made up of four constructs that build on one another:

\begin{itemize}
	\item \textbf{Verification-conditioned use} (Section~\ref{sec:rq1}): the individual decision rule, adopted independently of seniority, that governs whether a newcomer delegates a task to AI or performs it manually, based on the ability to check the result rather than on deadline or complexity.
	\item \textbf{Autonomy paradox} (Section~\ref{sec:rq3}): the simultaneous experience of feeling more capable and less in ownership of one's own output, which verification-conditioned use resolves in practice, through post-use review, rather than at the level of feeling.
	\item \textbf{First-person denial of dependence} (Section~\ref{sec:rq3}): the pattern of acknowledging AI over-reliance as a real risk for the profession in general while denying it applies to oneself, which reframes dependence as a collective risk under individual monitoring rather than a fixed trait.
	\item \textbf{Formative paradox} (this section): the higher-order tension that connects the other three constructs to the market context, in which the same AI-mediated learning that verification-conditioned use only partially offsets is occurring precisely as the market raises the value of the competence that learning puts at risk.
\end{itemize}

Read together, these four constructs offer a candidate explanatory model for how early-career software professionals experience generative AI: a moment-to-moment usage rule (verification-conditioned use) sits inside a self-perception tension (the autonomy paradox and its denial), both of which unfold inside a structural tension between individual learning and market demand (the formative paradox). None of the four constructs is specific to software development in its abstract form, which is one reason Section~\ref{sec:threats} argues for their transferability beyond the Brazilian software context studied here.

\section{Threats to Validity}
\label{sec:threats}

Following \citet{RH08}'s framework for software engineering case studies, four categories of threats apply.

\textbf{Construct validity.} Concepts such as ``autonomy'' or ``shallow learning'' could be interpreted differently by participants. This was mitigated by avoiding theoretical terms in the interview guide and by semantic-level inductive coding close to participants' own words. A residual threat remains: self-reported behavior may be more rationalized than actual practice, as the first-person denial of dependence (Section~\ref{sec:rq3}) illustrates, the study records what participants say about themselves, not necessarily what they do when unobserved.

\textbf{Internal validity.} Researcher bias in coding and interpretation is the main risk in thematic analysis. This was mitigated by strictly following the six phases of \citet{BC06}, anchoring every theme in direct quotations, and preserving divergent cases (e.g., E08, E13) rather than selecting only confirmatory accounts. A specific competing explanation deserves note for RQ4: the perceived rise in the entry barrier coincides with a broader contraction in entry-level technology hiring driven by macroeconomic factors unrelated to generative AI, and this study's design cannot isolate AI's share of that effect from the general market contraction.

\textbf{External validity.} The thirteen participants are interns and junior developers active in Brazil, mostly concentrated in the Pernambuco region, recruited through direct invitation and snowball sampling, which limits statistical generalization. Findings should be read as transferable rather than representative, reflecting the Brazilian context of 2025--2026. The underlying concepts (verification-conditioned use, the formative paradox) plausibly apply to other settings with a similar combination of an uncertain-quality tool, a market in transition, and professionals still in training, but this transferability claim is analytic, not empirically tested outside Brazil.

\textbf{Reliability.} Analysis was conducted by a single researcher, which is a limitation relative to studies that use independent double-coding with inter-coder agreement statistics. This was partially mitigated by coding the data in two separate passes, the first author's nine interviews, then the second interviewer's four, reapplying the same codebook without creating new codes, and by detailed documentation of the interview guide and coding process to support external audit.

Finally, this study was conducted during a period of fast-moving change in generative AI tools; the specific findings may lose descriptive validity as the market evolves, though the structural tensions identified, in particular the formative paradox, are expected to be more stable than the specific tool configurations that produce them.

\section{Conclusion}
\label{sec:conclusion}

This study investigated how generative AI use by interns and junior developers impacts their day-to-day work, learning process, task execution, and professional perception when entering the software job market, through thirteen semi-structured interviews analyzed with Braun and Clarke's six-phase thematic analysis \citep{BC06}. Across the four research questions, the main decision criterion for AI use is not deadline or complexity but the ability to check the result (\emph{verification-conditioned use}), materialized in a clear task split (CRUD, front-end, syntax, and testing delegated; architecture, business rules, and broad context retained). AI reconfigures learning by replacing documentation and forums as the first route, reducing depth, and removing the cognitive friction that sustained retention, while shifting senior consultations from ``how do I do this'' to ``is this correct?'' It increases perceived autonomy in mastered tasks while generating doubt about how much of that autonomy is one's own; trust comes from post-use review rather than from the output; and dependence is acknowledged collectively but denied in the first person, an arrangement sustained by four self-regulation strategies: reviewing before accepting, refusing to use AI without understanding it, asking the tool for explanations, and maintaining parallel unassisted practice. Finally, the entry barrier has risen as typical junior tasks migrate to AI, the professional differentiator has moved from writing to evaluating code, and the professional pyramid is reorganizing, with three coexisting postures toward the future: optimistic, concerned, and pragmatic.

The study's main theoretical contribution is the \textbf{formative paradox}: the same shallow, low-friction learning that AI induces undermines the critical-judgment competence the market has begun to demand as a differentiator. This reframes \emph{over-reliance} not only as an individual, cognitive problem but as a tension between the newcomer and a market that defines what counts as valuable skill.

\subsection{Practical Implications}
For \textbf{academia}, this study addresses a scarcity of qualitative research that jointly covers usage patterns, learning, autonomy, and market perception from the perspective of newcomers themselves, offering the formative paradox as a candidate axis for future work and curriculum design in computing education. For \textbf{companies and onboarding programs}, findings suggest that self-regulation does not track career length, since newcomers with only a few months already apply mature review strategies; this indicates the competence can and should be deliberately taught during onboarding rather than assumed to develop naturally with tenure, for instance by making verification-conditioned use an explicit part of intern and junior training rather than an implicit expectation. For \textbf{mentoring programs}, the shift in the type of question newcomers bring to seniors (Section~\ref{sec:rq2}) suggests mentors may need to proactively probe basic implementation understanding that no longer surfaces on its own, since AI has absorbed the questions that used to reveal gaps. For \textbf{educational institutions}, the formative paradox implies that protecting critical-judgment development may require deliberately preserving some AI-free ``productive struggle'' in coursework, even as AI fluency is also taught. For \textbf{newcomers themselves}, the study offers a peer-grounded diagnosis, verification-conditioned use, the autonomy paradox, self-regulation strategies, and the formative paradox, to help inform individual decisions about how, when, and how much to integrate AI into one's own practice.

\subsection{Future Work}
Five directions extend beyond the scope of this study: (1) designing formative processes, such as paired review, AI-free exercises, and explicit mentoring, that protect critical-judgment development without sacrificing AI-driven productivity; (2) explicit support for building self-regulation at earlier career stages, given that it does not appear to depend on tenure; (3) study of the professional pyramid's medium-term reorganization, as seniors orchestrate agents and non-developers enter technical territory; (4) longitudinal study of the same participants, to observe which self-regulation strategies survive productivity pressure and how verification-conditioned use evolves with seniority; and (5) triangulation with other data sources, such as workplace observation, commit and prompt analysis, and quantitative skill-retention instruments, particularly to investigate the first-person denial of dependence by comparing what participants say about their own AI use with what they actually do.

What the thirteen accounts make clear is that the underlying problem is not AI itself, but the relationship each newcomer builds with it, within a context that pressures for speed and is still learning to recognize the formative paradox. Self-regulation, as described by participants themselves, is today's available answer; academia, industry, and educational institutions can help ensure that answer does not rest solely on the newcomer, but becomes part of how the next generation of software professionals is trained.

The challenge, then, is no longer whether junior developers should use AI, that question is already settled in practice, but how they can develop the verification skills required to use it responsibly. Ironically, excessive reliance on AI may undermine the very competence that the future of software engineering increasingly demands.

\bibliographystyle{unsrtnat}
\bibliography{references}

\end{document}